\documentclass{aa}
\usepackage{graphicx}
\begin{document}

\title{The problem of the Pleiades distance}

   \subtitle{Constraints from Str\"omgren photometry of nearby field stars}

   \author{D. Stello
          \and
          P. E. Nissen
          }

   \offprints{D. Stello}

   \institute{Institute of Physics and Astronomy (IFA), University of Aarhus,
              DK-8000 Aarhus, Denmark\\
              \email{stello@ifa.au.dk, pen@ifa.au.dk}
             } 
   \date{Received April 4, 2001; accepted May 10, 2001}

\abstract{
The discrepancy between the Pleiades cluster distance based on
Hipparcos parallaxes and main sequence fitting is investigated on the
basis of Str\"omgren photometry of F-type stars. Field stars with the
same metallicity as the Pleiades have been selected from the $m_{1}$
index and a technique has been developed to locate the ZAMS of
these field stars in color-magnitude diagrams based on the
color/temperature indices
$b-y$, $v-y$, and $\beta$. Fitting the Pleiades to these ZAMS
relations results in a distance modulus of 5.61$\pm0.03$ mag in contrast to
the Hipparcos modulus of 5.36$\pm0.06$ mag. Hence, we cannot confirm the
recent claim by Grenon (\cite{grenon1999}) that the distance problem
is solved by adopting a low metallicity of the Pleiades
($[\mathrm{Fe/H}]_{\mathrm{Pleiades}}=-0.11$) as determined from Geneva photometry.
The metallicity sensitivity of the ZAMS determined by the field stars is
investigated, and by combining this sensitivity in all three
color/temperature indices $b-y$, $v-y$, and $\beta$ we get a
independent test of the Pleiades
distance modulus which support our value of 5.61 mag. 
Furthermore, the field star sample used for the comparison is tested
against theoretical isochrones of different ages to show that 
evolutionary effects in the field star sample are not biasing
our distance modulus estimate significantly.
Possible explanations of the Pleiades distance problem are discussed
and it is suggested that the discrepancy in the derived moduli may be
linked to a non-spherical shape of the cluster.
}

\maketitle

\section{Introduction}\label{introduction}
The Hipparcos Space Astrometry Mission has provided accurate 
absolute trigonometric parallaxes for roughly 120,000 
stars, which are distributed all over the sky, and hence relatively
accurate distance measurements for 
stars at a much larger distance than previous obtainable 
from ground based observations. This has given the 
opportunity to compare distances of several open clusters 
derived from direct trigonometric 
measurements with those derived from main sequence (MS) fitting.
For some of the clusters there are discrepancies between 
the derived distances, but in most cases the differences are within
the estimated uncertainties. An exception is the Pleiades for which the
distance modulus derived using the mean of the Hipparcos parallaxes is
almost 0.3 mag smaller than that derived using the MS fitting
technique.
Possible explanations of this anomaly are:
\begin{enumerate}
 \item The errors of the MS fitting technique may 
       be underestimated. This could arise from difficulties in 
       the technique itself, or it could be due to serious errors 
       in the adopted chemical composition of the Pleiades cluster.
 \item There may be systematic errors on small angular scales 
       in the sky of the Hipparcos parallaxes which are 
       underestimated. This could bias the inferred distance to 
       clusters that only cover a small angular area in the sky.
\end{enumerate}
If these possibilities can be excluded we may have to draw the important
conclusion that the theory of stellar structure and evolution, is
incomplete or in other words: The Vogt-Russell theorem 
that the location of a star in the Hertzsprung-Russell
diagram is uniquely determined by its mass, age, and composition is
violated. It is this aspect of the Pleiades distance problem that makes
it so interesting.

The mean parallax of the Pleiades cluster inferred from the Hipparcos
data ranges from 8.60$\pm0.24$ mas (Mermilliod et al. \cite{merm97}) to
8.45$\pm0.25$ mas (van Leeuwen \cite{vanleeuwen99}). These parallaxes
correspond to a
distance interval of 116$\pm3$ pc to 118$\pm4$ pc or a distance modulus
interval of 5.33$\pm0.06$ mag to 5.37$\pm0.07$ mag. 
These distance moduli should
be compared with those found from the MS fitting method. 
Pinsonneault et al. (\cite{Pin98}) find a distance modulus of 5.60$\pm0.05$
mag, based on an
extensive multi color MS fitting analysis. They
make use of several open clusters to check for different possible
error sources, and both isochrones and an empirical Hyades MS are
used as the zero point of the ZAMS. 
Pinsonneault et al. (\cite{Pin98}) suggest that the discrepancy 
between the results from Hipparcos and the MS fitting method is due
to spatial systematic errors on small angular scales in the Hipparcos
data (Pinsonneault et al. \cite{Pin98}, Fig. 20) which are larger
than expected (Lindegren \cite{lindegren1988}, \cite{lindegren1989},
\cite{lindegren1997}).
From a comparison of the Pleiades MS with those of the Hyades and
$\alpha$ Persei clusters Eggen (\cite{eggen1998}) also concludes that
the Hipparcos parallax distance of the Pleiades may be in error by
some 10 \%.
An investigation of the possible spatial systematic errors in the
Hipparcos data is performed by Narayanan \& Gould (\cite{narayanan99})
who confirm 
the distance modulus of around 5.6 mag by an estimate of 5.58 mag,
though with a fairly large error of $\pm0.18$ mag. Their investigation
is based on a variant of the moving cluster method to get 
distances for each individual cluster member. 
On the basis of the stated spatial systematic errors in the Hipparcos
data Robichon et al. (\cite{Rob99}) make an extensive 
investigation of this problem, which makes use of the method of analyzing Hipparcos
intermediate data described by van Leeuwen \& Evans
(\cite{leeuvenevans1998}). They recalculate the Hipparcos
parallaxes and find a distance modulus of 5.36$\pm0.06$ mag for the
Pleiades which is the value adopted in this paper for comparisons.

Recently, Grenon (\cite{grenon1999}) has claimed that the Pleiades
distance problem is solved by adopting a cluster metallicity of
$[\mathrm{Fe/H}]_{\mathrm{Pleiades}}=-0.11$ as determined from Geneva 
photometry instead of
$[\mathrm{Fe/H}]_{\mathrm{Pleiades}}\simeq0.0$ as determined from high 
resolution spectroscopy
(Boesgaard \& Friel \cite{BoesFriel}). 
In particular, Geneva photometry points to a large difference in the
metallicity of Praesepe and Pleiades ($\Delta[\mathrm{Fe/H}]=0.28$). To a
large extent this explains the offset between the two clusters
in various color-magnitude diagrams (Robichon et al. \cite{Rob2000}).

The purpose of the current study is to investigate the discrepancy 
between the Pleiades distance inferred from the Hipparcos 
mean parallax and from the MS fitting method by comparing Str\"omgren
$uvby-\beta$ photometry of Pleiades F-type stars with field stars 
having the same metallicity as the Pleiades. 
We start out with a presentation of the Pleiades cluster and field star data
followed by the calibration formulas and selection of stars
used for the rest of the reductions. Then the MS fitting analysis is
described 
including the fitting technique, which has been developed to locate the
ZAMS of the field star sample, and how we test the age range of the field
stars. Furthermore, a metallicity sensitivity analysis is presented
and finally we discuss possible answers to the Pleiades problem
including the suggestion that the discrepancy in the
derived distance moduli could be a real effect caused by the structure
of the cluster.

\section{MS fitting with Str\"omgren photometry}\label{min_del}
The Pleiades, shifted to the distance inferred from 
the Hipparcos mean parallax, are compared to nearby 
field stars of the same metallicity in a 
color-magnitude diagram. This is done to see if the 
ZAMS defined by the field stars (hereafter denoted 
ZAMS$_{\mathrm{Field}}$) does coincide with the 
Pleiades MS. Further, the distance modulus of the 
Pleiades is inferred from a best fit by the 
comparison of the 
Pleiades MS and the ZAMS$_{\mathrm{Field}}$. 
This investigation also includes a sensitivity analysis 
of the ZAMS$_{\mathrm{Field}}$ locus to changes 
in [Fe/H]. 
The analysis is performed for F-type stars, 
and in three color/temperature indices 
$b-y$, $v-y$, and $\beta$, to check for 
robustness and possible hidden errors in 
the MS fitting technique.

The advantage of using field stars to define the zero 
point of the distance modulus is that the investigation 
will be purely empirical, and not affected by some 
possible lacking ingredients in the theoretical 
calculations of the isochrones. So this investigation only 
relies on the Vogt-Russell theorem (Sect. \ref{introduction}). 
The shape of the Pleiades MS turns out to be very 
similar to the ZAMS$_{\mathrm{Field}}$ shape and hence we avoid the 
problem of fitting to isochrones which do not match the 
cluster MS at every temperature interval. 
Using F-type stars gives the opportunity to correct 
for interstellar reddening and to estimate [Fe/H], based 
on the Str\"omgren indices and available empirical 
calibration formulas. 
The assumed shape of the ZAMS$_{\mathrm{Field}}$ used for the comparison
between the Pleiades MS and the field stars is found by a second-order
robust least squares fit to the Pleiades stars.

\subsection{Data} \label{data}
The $uvby-\beta$ data for the Pleiades is 
taken from Table II of Crawford \& Perry (\cite{crawperry76})
(hereafter CP76), which contains 
members classified as F-type stars by CP76. 
The mean error of one observation, was determined 
from the internal scatter in the measurements of all 
the Pleiades stars (Table I CP76), and were given as follows, 
$\sigma (\beta)=0.011$, $\sigma (b-y)=0.009$, 
$\sigma (m_{1})=0.011$, and $\sigma (c_{1})=0.012$ (CP76). 
Taking into account that each star has been observed about 5 times we
obtain the following typical mean errors of the mean indices of one
star: $\sigma (\beta)=0.005$, $\sigma (b-y)=0.004$, 
$\sigma (m_{1})=0.005$, and $\sigma (c_{1})=0.005$ mag. 

The magnitudes of the stars in CP76 
were adopted from Johnson \& Mitchell (\cite{johnmitc}) 
(based on the $UBV$ system), so to get the Str\"omgren 
values ($y$ transformed to $V$) the star magnitudes are 
taken from the WEBDA database (Mermilliod \cite{webda}). For about
half of the stars the Str\"omgren value is not 
available, and the magnitudes are 
taken to be the average of the $V$ values from $UBV$ 
observations (also given in the WEBDA database). 
It is noted that for stars with both $uvby$ and $UBV$ photometry, the
$V$ magnitudes agree within $\pm0.02$ mag.

The field star data has been taken from a catalogue of 
$\sim 30.000$ stars observed in $uvby-\beta$ (Olsen 
\cite{olsen99}). This catalogue has been made by 
merging five published catalogues; all by E.H.Olsen. 
The sample used in this investigation (F-type stars) 
is based on three of these catalogues (Olsen \cite{olsen83}, \cite{olsen88}, 
\cite{olsen94}). The overall RMS 
internal error of one observation is 
$\sigma (V)=0.005$, $\sigma (b-y)=0.004$, 
$\sigma (m_{1})=0.006$, $\sigma (c_{1})=0.007$, 
and $\sigma (\beta)=0.007$. These errors are 
the conservative ones, in some of the catalogues they 
were in fact $\sim 0.002$ mag smaller, but the 
conservative ones are adopted in this investigation. 
The majority of the stars were observed only once and 
a few 2 or 3 times, so the errors in the mean 
photometric values per star are not significantly different from the
errors stated above.

The absolute magnitudes are derived using 
the reddening corrected apparent magnitudes $V_{0}$ 
(transformed from $y$), and the 
Hipparcos parallaxes (ESA \cite{esa1997}). The mentioned possible
spatial systematic error in the Hipparcos parallaxes does not affect
the locus of the ZAMS$_{\mathrm{Field}}$, because 
it can only have effects on small angular scales, and 
the field stars are distributed ``randomly'' all 
over the sky. Thus the error contributions from the parallaxes in the 
zero point for the ZAMS$_{\mathrm{Field}}$ is the global 
parallax error, which is less 
than 0.1 mas (Arenou et al. \cite{arenou1997}). 

In addition to the internal errors for the Pleiades photometry by CP76
and for the field stars by Olsen (\cite{olsen99}) there could be
systematic differences between the two sources. Especially the $\beta$
index is critical, because the reddening and hence the corrected color
indices $(b-y)_{0}$ and $(v-y)_{0}$ is determined from $\beta$
(Sect. \ref{reddening}). It is, however, very unlikely that systematic
errors in $\beta$ could be large enough to explain the offset between
the Pleiades and the field stars, which is of the order of 0.02 mag in
$\beta$ if we adopt the Hipparcos distance modulus of the Pleiades. In
this connection we note that photometric observations of the $\beta$
index is quite straightforward; no extinction correction is needed and
the transformation to the standard system is linear without color
terms. As discussed by Olsen (\cite{olsen83}), systematic
differences in $\beta$ obtained with different telescopes and filter
sets are 0.005 mag at most. Furthermore, we note that two of the
Pleiades F-type stars from CP76 (Hz II 739 and 948) happen to be in
Olsen (\cite{olsen99}). The differences
$(\mathrm{Olsen}-\mathrm{CP76})$ are 0.007 and 0.008,
respectively. This does not point to any large systematic errors, and
a correction for this difference would in fact increase the offset of
the Pleiades with respect to the field stars, but of course we cannot
draw any strong conclusions from two stars only.

\subsection{Calibration and selection}
\subsubsection{Reddening} \label{reddening}
To correct for reddening, the color excess is 
determined for individual stars as 
$E(b-y)=(b-y)-(b-y)_{0}$, where $(b-y)_{0}$ is 
found by an iterative 
calculation based on the empirical calibration 
given by Crawford (\cite{crawford75a}): 
\begin{eqnarray}
(b-y)_{0} &=& 0.222 + 1.11 \Delta \beta + 2.7 \Delta \beta^{2} - \\
          & & 0.05 \delta c_{0} - (0.1 + 3.6 \Delta \beta) \delta m_{0}
              \; , \nonumber
\end{eqnarray}
\noindent where
\begin{equation}
\delta c_{0} = c_{0} - c_{0,\mathrm{ZAMS}}(\beta) \; ,\  
\delta m_{0} = m_{0,\mathrm{Hyades}}(\beta) - m_{0} \; ,
\label{c0andm0}
\end{equation}
\begin{equation}
c_{0} = c_{1} - 0.20 \cdot E(b-y) \; ,\  
m_{0} = m_{1} + 0.30 \cdot E(b-y) \; ,
\end{equation}
\noindent and $\Delta \beta = 2.72 - \beta$. The 
standard relation between $\beta$, $m_{0,\mathrm{Hyades}}$, 
and $c_{0,\mathrm{ZAMS}}$ is found by interpolation between 
the data points 
given in Table I by Crawford (\cite{crawford75a}). 
The iterative procedure uses the four Str\"omgren indices for each 
star, and an initial 
guess for the color excess as input data. The output 
(individual color excesses) was obtained when $E(b-y)$ 
converged at the 0.0001 level. 
The expected error for $E(b-y)$ is found as 
\begin{equation}
\sigma[E(b-y)] = \big[\sigma^{2}(b-y) + 
\sigma^{2}[(b-y)_{0}]\big]^{1/2} \; , 
\end{equation}
\noindent where
\begin{eqnarray}
\sigma^{2}[(b-y)_{0}] &=& \sigma ^{2}(\Delta\beta)
      (1.11 + 5.4\Delta\beta + 3.6\delta m_{0})^{2} + \\
      &  &\sigma ^{2}(\delta m_{0})(0.1+3.6 \Delta\beta)^{2} + 
      0.05^2\sigma ^{2}(\delta c_{0}) , \nonumber
\label{byerror}
\end{eqnarray}
\noindent with $\sigma(\Delta\beta)=\sigma(\beta)$, 
$\sigma(\delta m_{0}) \simeq \sigma(m_{1})$, and 
$\sigma(\delta c_{0}) \simeq \sigma(c_{1})$.  
An estimate of this error is obtained by inserting 
the mean values of 
$\Delta\beta$ and $\delta m_{0}$ from the samples. 
For the Pleiades stars it is 
$\sigma [E(b-y)]_{\mathrm{Pleiades}} = 0.008$ mag, while it 
for the field stars is 
$\sigma [E(b-y)]_{\mathrm{Field}} = 0.010$ mag\label{ebyerror}.

\subsubsection{[Fe/H] calibration}
To be able to select nearby field stars with the 
same metallicity as the Pleiades, [Fe/H] 
is calculated for the Pleiades using the 
empirical calibrations of Nissen (\cite{nissen81}). 
The calibration formula is as follows:
\begin{equation}
\textrm{[Fe/H]} = -[10.5 + 50(\beta - 2.626)] 
\delta m_{0} + 0.12 \label{fehformula}
\end{equation}
\noindent where $\delta m_{0}$ is like in Eq. 
(\ref{c0andm0}), and the constant of 0.12 is the 
adopted $[\mathrm{Fe/H}]_{\mathrm{Hyades}}$. 
The adopted $[\mathrm{Fe/H}]_{\mathrm{Hyades}}$ does not 
affect the result of the comparison of the Pleiades 
with the field stars, because it is only relative. But 
the estimate of the absolute $[\mathrm{Fe/H}]_{\mathrm{Pleiades}}$ value 
is of course dependent on the assumed 
$[\mathrm{Fe/H}]_{\mathrm{Hyades}}$. The expected error in this calibration 
is obtained by a procedure similar to that presented 
in Sect. \ref{reddening}, but this time 
$\sigma(\delta m_{0})=\sigma(m_{0})$ is used. The results are 
$\sigma ([\mathrm{Fe/H}])_{\mathrm{Pleiades}}=0.07$ and $\sigma 
([\mathrm{Fe/H}])_{\mathrm{Field}}=0.07$.

\subsubsection{Reduction step by step}
The criterion for 
being an F-type star is set to be $2.59<\beta<2.72$, 
which is the $\beta$ range used by Crawford (\cite{crawford75a}) and 
Nissen (\cite{nissen81}) for their calibration formulas (reddening 
and metallicity). 

To keep as many Pleiades members as possible, no 
stars from Table II (CP76) are 
rejected as a start except Hz II 948 which appear to be a non-member in
both CP76 and the WEBDA database.
For each member star the reddening is calculated 
by the procedure described in Sect. \ref{reddening}, 
and their photometry measurements are individually 
corrected. 
The mean reddening obtained is 
$\langle E(b-y) \rangle_{\mathrm{Pleiades}} = 0.031\pm0.004$ 
mag\footnote{This value is based on all stars from Table II of CP76
  except the likely non-members Hz II 739 and Hz II 948 (see Sect. \ref{ms_fitting}).}, 
and the star-to-star RMS scatter is 
$S[E(b-y)]_{\mathrm{Pleiades}} = 0.022$ mag. Compared 
with the expected  error of the 
$E(b-y)$ determination ($\sigma [E(b-y)]_{\mathrm{Pleiades}} = 
0.008$ mag), this indicates significant star-to-star reddening 
differences across the cluster. 
The mean color excess for the Pleiades obtained here 
is in quite good agreement with former 
results obtained from other investigations, which in 
general are in the range of about 0.03-0.04 mag (e.g. 
Pinsonneault et al. \cite{Pin98} used 
$E(b-y)=0.7\times0.04\textrm{ mag }\simeq0.03$
mag; here using the relation between $E(b-y)$ and $E(B-V)$ from
Crawford \cite{crawford75b}). 

The reddening corrected $m_{0}$ values together with 
the $\beta$ observations are then used as input 
in Eq. (\ref{fehformula}), to get the Pleiades 
metallicity. The mean value derived is 
$\langle [\mathrm{Fe/H}] \rangle_{\mathrm{Pleiades}} = 0.01\pm0.02^1$.  
This value is in very good agreement with spectroscopic results, which  
mostly come out with a near solar metallicity for 
the Pleiades (e.g. Boesgaard \& Friel \cite{BoesFriel}). 
The star-to-star RMS scatter is 
$S([\mathrm{Fe/H]})_{\mathrm{Pleiades}} = 0.13$, 
which is somewhat larger than the expected error
($\sigma([\mathrm{Fe/H}])_{\mathrm{Pleiades}}=0.07$). 

From the catalogue of field stars, used 
in this investigation, there are 12658 stars which have 
$2.59<\beta<2.72$ (thus F-type stars), but 1194 stars of this 
group do not have Hipparcos parallaxes, so the 
absolute magnitude could not be derived, and they are 
therefore rejected. For each star in the remaining sample 
the reddening is calculated, as described in Sect. 
\ref{reddening}, and the photometry of every star is 
individually corrected. Due to statistical fluctuations and a low mean
reddening of the field star sample ($\langle E(b-y)\rangle =0.009$)
some stars turn
out to have slightly negative values of $E(b-y)$. In order to avoid
any bias these negative values were not changed.
Finally, the reddening corrected $m_{0}$ values and the $\beta$ 
observations can be used as input data in Eq. 
(\ref{fehformula}), to derive the metallicity 
for every star. 

We choose a metallicity range of  
$-0.10<[\mathrm{Fe/H}]<0.12$ which is comparable to the Pleiades mean
metallicity plus/minus a representative estimate of the metallicity
scatter, and the number of F-type 
field stars remaining in this interval is 3389. 
The mean [Fe/H] of the remaining sample is not equal 
to $\langle [\mathrm{Fe/H}] \rangle_{\mathrm{Pleiades}}$ because the 
metallicity profile of the original field star sample 
peaks around $[\mathrm{Fe/H}]=-0.15$, thus [Fe/H] for the 
remaining sample is slightly shifted (by 0.01 dex) 
to a lower [Fe/H]. But since this is only half the error of the mean
of the Pleiades [Fe/H] the effect is ignored. 

An additional selection of the field stars is made 
on the basis of the relative parallax error. If the 
$M_{V_{0}}$ vs. $(b-y)_{0}$ diagram is considered, 
the error in the absolute magnitude $M_{V_{0}}$ 
of the field stars is affected by the errors in 
$V_{0}$, $(b-y)_{0}$, and 
the distance modulus (through $\pi$, the parallax). 
From the errors given in Sect. \ref{data} and 
\ref{reddening}, the error in $V_{0}$ 
can be estimated as: 
$\sigma (V_{0}) = [\sigma^{2} (V) + (4.3 \, \sigma 
[E(b-y)]_{\mathrm{Field}})^{2}]^{1/2} = 0.044$ mag. 
The size of the effect on $\sigma(M_{V_{0}})$ from 
$\sigma [(b-y)_{0}]$ depends on the slope of the 
ZAMS in the color region of interest. A test plot 
is made to find the approximate slope, and it is 
found to be $\sim 12$. 
The effect from $\sigma (\pi)/\pi$, on the distance 
modulus, is found by differentiating the relation 
$(m-M)=5\log(\frac{1}{\pi})-5$, with respect to 
$\pi^{-1}$, where $(m-M)$ is the distance modulus. 
This now leads to the following expression: 
\begin{equation}
\sigma^{2} (M_{V_{0}}) \simeq \sigma^{2}(V_{0})+(12 \, \sigma 
[(b-y)_{0}])^{2}+(2.17 \, \sigma (\pi)/\pi)^{2} \; . 
\label{sigma_mag_formular}
\end{equation}
The optimized choice of the upper limit of 
$\sigma (\pi)/\pi$, is when $2.17 \, \sigma (\pi)/\pi 
\sim$ max$\{\sigma(V_{0})$, $12 \, \sigma [(b-y)_{0}]\}
\sim 0.12$ mag  
(where $\sigma [(b-y)_{0}]_{\mathrm{Field}}=0.010$ mag; Eq. 
(\ref{byerror})), 
which suggests $\sigma (\pi)/\pi \sim 0.05$. 
If a similar consideration is made in 
the $M_{V_{0}}$ vs. $(v-y)_{0}$ or $M_{V_{0}}$ vs. 
$\beta$ diagram, the suggestion would be  
$\sigma (\pi)/\pi \sim 0.04$. 
To avoid different 
samples of field stars in the three investigation parts, 
parallax measurements to a 
5\% accuracy are chosen, which left a sample of 782 stars. This 
seems to be the most reasonable choice, since the 
sample is large enough to make a clear definition of the ZAMS, 
and a lower error always is desirable.

\subsection{MS fitting analysis}\label{ms_fitting}
The MS fitting analysis is carried out in three diagrams 
$M_{V_{0}}$ vs. $(b-y)_{0}$, $M_{V_{0}}$ vs. $(v-y)_{0}$, 
and $M_{V_{0}}$ vs. $\beta$. Often only the
$(b-y)_{0}$ case will be illustrated in the figures, but the results of
the other diagrams will be given. Though $\beta$ is not a 
color, all three diagrams will in the following be 
denoted color-magnitude diagrams, and the 
specification ``F-type'' stars will be omitted, thus 
``the stars'' or ``all the stars'' simply refers to the 
sample of F-type stars used in this investigation. 
A test plot of the Pleiades showed that there was a 
outlying star which was significantly cooler than the rest 
of the sample, and $\sim 1$ mag 
above the Pleiades MS (it has spectral type G0; 
Mendoza \cite{mendoza56}). The star is 
rejected as a likely non-member. The Pleiades mean color excess and
metallicity did not change significantly if the G0 star was
included in the sample or not. The changes were only 0.002 mag in
the average color excess, and $\langle[\mathrm{Fe/H}]\rangle$ changed
by 0.003 dex.

With the errors stated in Sect. \ref{data} and 
Sect. \ref{reddening} the estimated 
error per one star in $M_{V_{0}}$ for the field stars 
[Eq. (\ref{sigma_mag_formular})], is 
in the range of 0.14-0.17 mag\label{sigma_mag_field} 
for all three examined color-magnitude diagrams 
(smallest for the $(v-y)_{0}$ diagram, 
and largest for the $(b-y)_{0}$ and $\beta$ diagrams, which is expected
since the ZAMS in the $(v-y)_{0}$ diagram is less steep than in the
other two diagrams). 

\begin{figure*}\centering
\includegraphics{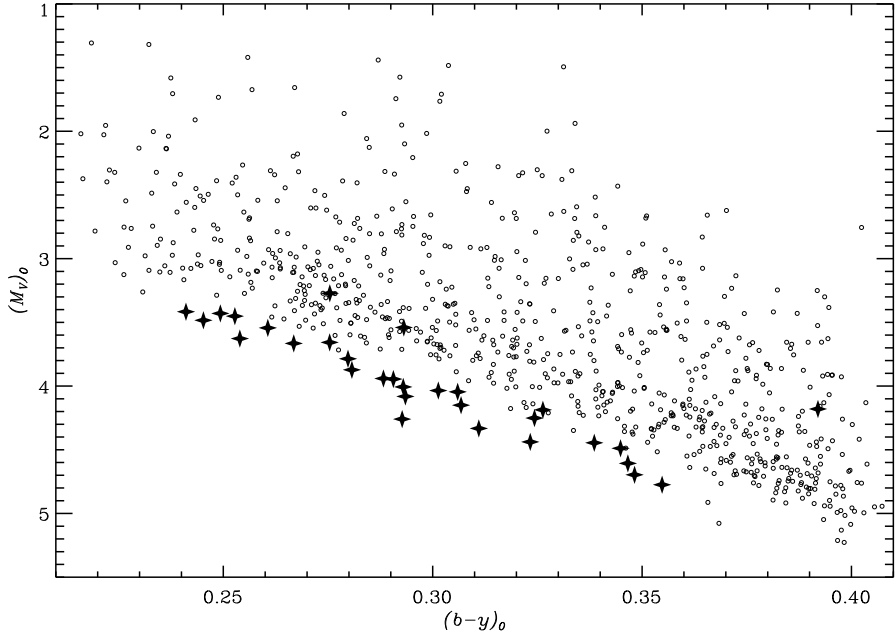}
\caption[Field and Pleiades stars shifted to the 
Hipparcos distance]{\label{plei_hip_dist}
{Color-magnitude diagram of the selected field stars with 
$-0.10<[\mathrm{Fe/H}]<0.12$ and $\sigma (\pi)/\pi<0.05$ (empty circles), 
compared to the Pleiades (filled stars) adopting the Hipparcos
inferred distance modulus of 5.36 mag (from Robichon et al. \cite{Rob99}). 
The Pleiades star at the cooler end of the diagram
is a G0 star and most likely a non-member, and the two stars (Hz II
1338 and 1912) about 0.6
mag brighter than the mean relation defined by the others Pleiades
stars are binaries. 
}}
\end{figure*}

In Fig. \ref{plei_hip_dist} the field stars 
are plotted together with the Pleiades stars adopting the Hipparcos
mean parallax of the cluster. 
This plot clearly shows the discrepancy between the 
locus of the field stars and that of the Pleiades. 
The plots in all three color-magnitude diagrams look 
quite similar. In addition to the deviating G0 star 
there are two Pleiades stars (Hz II 1338 and 1912) which are 
$\sim0.6$ mag brighter than the mean relation defined by the others. 
According to Mermilliod et al. (\cite{merm92}), one of the
stars is a spectroscopic binary, and the other a visual binary, in
agreement with their shift in brightness.

\subsubsection{The stellar magnitude distribution}\label{star_dens_profil}
To get an estimate of the distance modulus of the Pleiades 
we have to analyze the magnitude distribution, relative to the
Pleiades MS, of the field stars 
which define our zero point of the ZAMS locus. 
The observed magnitude distribution is a convolution of the underlying
evolutionary 
distribution of the field star sample and a Gaussian profile 
coming from the observational errors in $M_{V_{0}}$. 

A robust least squares fit (second-order) (Freudenreich
\cite{Freudenreich}) of the Pleiades 
MS is made in the $V_{0}$ vs. color diagrams.
The derived polynomial of the Pleiades MS is 
used to define the shape of the ZAMS$_{\mathrm{Field}}$. To find the
distance modulus of the Pleiades the polynomial is 
shifted by the magnitude which make it fit to 
the ZAMS$_{\mathrm{Field}}$. The advantage 
of this method is that the robust least squares fit of the 
Pleiades MS will not be significantly affected by the stars which 
lie far from the Pleiades MS. This means that e.g. double 
stars will not bias the locus of the Pleiades MS significantly. 
The disadvantage is that the shifted Pleiades MS fit 
may not match the ZAMS$_{\mathrm{Field}}$ 
perfectly, because the shape of the fitted Pleiades 
MS is sensitive to the small sample of 
Pleiades data points (29 stars). 
The shifting of the Pleiades MS is done by calculating the individual
distance moduli of every field star with respect to 
the polynomial fit of the Pleiades MS. This give the magnitude
distribution of the field stars relative to the Pleiades MS. 
All the distance moduli
are evaluated as input in the likelihood function which is the simultaneous
probability function of all data points. The probabilities of the
individual data points are described by the result
of the convolution mentioned above. The underlying evolutionary 
distribution is approximated by $\propto \exp(-x/\tau)$ with a sharp edge 
at the ZAMS$_{\mathrm{Field}}$ locus, where $\tau$ is
the fall-off rate due to evolution and binarity of the star
sample. The width of the Gaussian is characterized by
the observational error $\sigma(M_{V_{0}})$. 
The mathematical expression of the convoluted function is:
\begin{eqnarray}
f(u) &=& \frac{p_{1}}{p_{2}}
         \exp( \frac{p_{4}^{2}}{2p_{2}^{2}} - \frac{u-p_{3}}{p_{2}} )
         \frac{1}{\sqrt{2\pi}p_{4}}
         \int_{-\infty}^{y} \!\!\!\!
         \exp( -\frac{v^{2}}{2p_{4}^{2}} )dv \nonumber \\
     &=& \frac{p_{1}}{2p_{2}}
         \exp( \frac{p_{4}^{2}}{2p_{2}^{2}} - \frac{u-p_{3}}{p_{2}} )
         ( 1+\mathrm{Errf}( \frac{y}{\sqrt{2}p_{4}} ) ) 
\label{convolution}
\end{eqnarray}
where $y=u-p_{3}-p_{4}^{2}/p_{2}$, $p_{1}$ is the normalization constant,
$p_{2}=\tau$ , $p_{3}$ is the ZAMS$_{\mathrm{Field}}$ locus, 
$p_{4}=\sigma(M_{V_{0}})$, and Errf(t) is the IDL error function.
A best fit is obtained when the likelihood function takes its maximum value
which we find by changing $\tau$, $\sigma(M_{V_{0}})$, and the
locus of the sharp edge of the ZAMS$_{\mathrm{Field}}$ as free parameters.

\begin{figure}\centering
\includegraphics{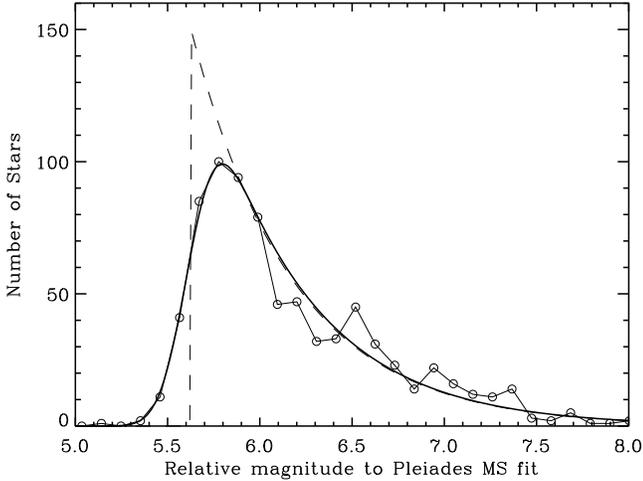}
\caption[Field Star Density Profile]{\label{dens_profile}
{The field star magnitude distribution, with 
respect to the polynomial fit of the Pleiades MS, in the 
$M_{V_{0}}$ vs. $(b-y)_{0}$ diagram. Empty circles shows the individual
distance moduli binned, with a bin size of 0.1 mag and connected with
the thin line. Dashed line 
indicates the assumed underlying star distribution without observational
errors defining the point of the ZAMS$_{\mathrm{Field}}$. The thick line 
is the best fit to the data of the convolution of the underlying 
star distribution with a Gaussian profile which corresponds to the
observational error.
}}
\end{figure}

The underlying evolutionary distribution (dashed line) together with
the convoluted function of best fit (thick line) is shown in Fig. 
\ref{dens_profile}. 
Additionally, the individual distance moduli are binned and 
over plotted to illustrate the field star magnitude distribution for
comparison (thin line and empty circles). The point of the
ZAMS$_{\mathrm{Field}}$, is the vertical dashed line. 

The fitted $\sigma(M_{V_{0}})_{\mathrm{Field}}$ for all three color-magnitude diagrams 
end up in the range 0.12-0.13 mag which is near the expected errors
derived in Sect. \ref{ms_fitting} on page \pageref{sigma_mag_field}.
It indicates that the shape of the Pleiades 
polynomial match the ZAMS$_{\mathrm{Field}}$ shape very well in all three 
color-magnitude diagrams. 
The Pleiades distance modulus found by the fits are: $5.62\pm0.02$
mag, $5.61\pm0.02$ mag, and $5.61\pm0.03$ mag in the $(b-y)_{0}$, $(v-y)_{0}$, and
$\beta$ diagram respectively. 
The stated errors are the quadrature sum of two errors. The first coming
from the uncertainty in the vertical positions of the Pleiades MS fits,
which is the star to star scatter around the Pleiades MS fit divided
by the square root of the number of stars 
($S_{V_{0},\mathrm{Pleiades}}(\textrm{Pleiades MS
  fit})/\sqrt{\textrm{\#stars}}$). For the three color-magnitude
diagrams these errors are: 0.019 mag, 0.017 mag, and 0.031 mag in the
$(b-y)_{0}$, $(v-y)_{0}$, and $\beta$ diagram respectively\footnote{These
  errors are from slightly around one to 3/2 times larger than the
  pure observational scatter in $V_{0,\mathrm{Pleiades}}$, and show different real
  effects like binarity, which introduces extra scatter around the
  perfect ZAMS locus.}.
The second error is found as the points
where the log likelihood function has fallen by 0.5 from its 
maximum value by changing the Pleiades distance modulus step by step 
around its optimum value, and optimizing the two other parameters for 
every step. The $\pm 1\sigma$ interval is approximated by a
symmetric interval around the maximum value by a parabola fit of the 
log likelihood function. This approximation is less than 
the resolution. For the three color-magnitude
diagrams these errors are: 0.014 mag, 0.015 mag, and 0.015 mag in the
$(b-y)_{0}$, $(v-y)_{0}$, and $\beta$ diagram respectively.

It is important that the left tail of the field star's magnitude
distribution is fitted well because this is
the region where the ZAMS$_{\mathrm{Field}}$ has to be found. The extremely 
good fit at the left tail of Fig. \ref{dens_profile} 
supports the trustfulness of the method used in this investigation. 
The right tail is 
more affected by the evolutionary and binary distribution of the field 
star sample, and it could be argued that the assumed exponential
fall-off at the right part of the profile is rather simplified, but
changing it would not affect the fitted ZAMS$_{\mathrm{Field}}$ locus
significantly because this part of the diagram is separated by 
several standard errors from the ZAMS$_{\mathrm{Field}}$ locus.

The method of finding the distance modulus presented here seems very 
robust because the distance moduli, errors in the distance moduli 
and $\sigma(M_{V_{0}})$ are consistent with one another and their
estimates in the three color-magnitude diagrams. 
By changing the underlying evolutionary distribution it is seen that 
the estimated distance moduli are quite stable. 
We consider the following different underlying star
distributions. Assume the underlying fall-off to be in two
steps. \label{two-step}First a rather steep fall-off followed by a less steep
fall-off. This scenario would fit the Gaussian to be wider (but could
still be consistent with the estimates given in
Sect. \ref{sigma_mag_field} on page
\pageref{sigma_mag_field}), and the Pleiades distance modulus would be
fitted to be slightly larger (ZAMS$_{\mathrm{Field}}$ closer to observed
maximum density). 
The only way to get a 
lower estimated distance modulus is by assuming a slower fall-off for 
the right tail of the underlying distribution; the extreme being a 
wide box function with one edge at the ZAMS$_{\mathrm{Field}}$. Under this 
assumption the ZAMS$_{\mathrm{Field}}$ will be at the point of half maximum 
of the observed distribution, which in this case means a lower 
distance modulus by less than 0.05 mag, but at the cost of an
unacceptable bad fit to the data. 
One can of course get an even lower estimate if it is 
assumed that the underlying star density will increase on the right
side of the 
ZAMS$_{\mathrm{Field}}$ in Fig. \ref{dens_profile}, which means that 
the maximum density of the underlying distribution is above the 
ZAMS$_{\mathrm{Field}}$. But the extremely
good fits at the left tail (Fig. \ref{dens_profile}) and the
consistent determinations of the Gaussian widths tell us that the
underlying star distribution must have a sharp edge as indicated in
Figure \ref{dens_profile}.
To see if the field star sample does indeed contain stars not evolved
significantly away from the ZAMS, the 
sample is compared with a series of isochrones in Sect. \ref{agetest},
and the assumed underlying evolutionary distribution is tested by the
aid of evolutionary tracks.

\begin{figure*}\centering
\includegraphics{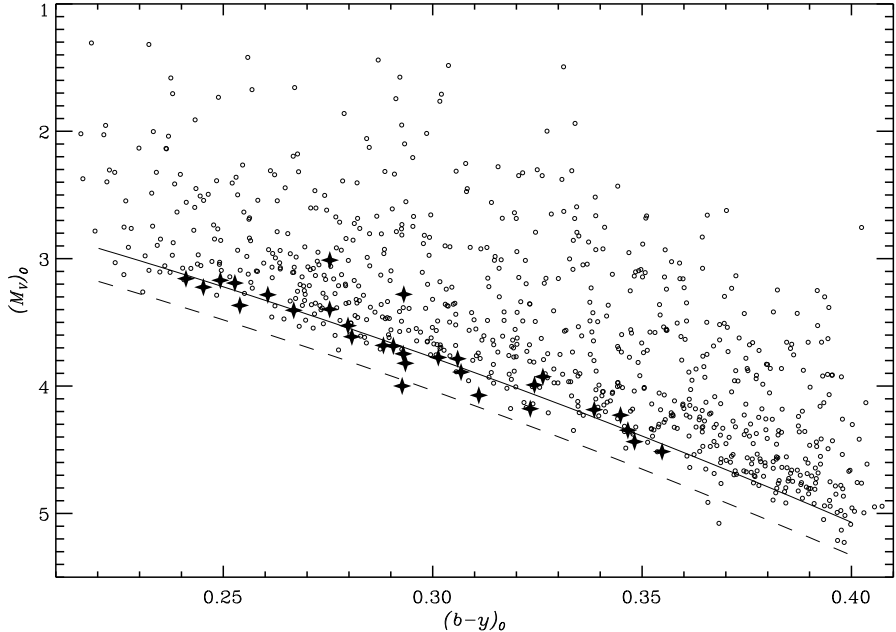}
\caption[Pleiades MS fit shifted to the ZAMS$_{\mathrm{Field}}$]
{\label{plei_my_dist}
{Color-magnitude diagram of the selected field stars with 
$-0.10<[\mathrm{Fe/H}]<0.12$ and $\sigma (\pi)/\pi<0.05$ (empty circles), 
compared to the Pleiades (filled stars) (excluding the likely
non-member) 
adopting the distance modulus of 5.62 mag, which make the 
Pleiades MS match the ZAMS$_{\mathrm{Field}}$. 
The solid line is the polynomial fit to the Pleiades MS, and the
dashed line indicates the location of the Pleiades
MS according to the Hipparcos distance. 
}}
\end{figure*}

Figure \ref{plei_my_dist} shows a plot similar to Fig. \ref{plei_hip_dist}, 
but instead of the adopted Pleiades distance modulus of Hipparcos 
the estimated value of 5.62 mag, which make the fitted Pleiades MS 
polynomial match the ZAMS$_{\mathrm{Field}}$, has been used. Furthermore, 
the polynomial fit to the Pleiades MS is shown, shifted by 5.62 mag
(solid line) and by 5.36 mag (dashed line). It is easy to see that
the Pleiades shifted by 5.62 mag gives a much better fit to the field
stars compared to the Hipparcos value.

\subsubsection{The age range}\label{agetest} 
The age of the field star sample which is used to determine the 
Pleiades distance modulus is analyzed by plotting isochrones 
of different ages together with the field stars. 

The isochrones used are taken from Lejeune \& Schaerer (\cite{lejeune}), 
those they denote as ``basic grid'' with solar 
metallicity. The effective temperature coming from the isochrones 
is transformed to the $(b-y)_{0}$ color index by the calibration 
of Alonso et al. (\cite{alonso}) using a mean value of the $c_{1}$
index in their Eq. (9). The possible systematic error in
$(b-y)_{0}$ from this transformation is of the order of 0.02 mag.

\begin{figure}\centering
\includegraphics{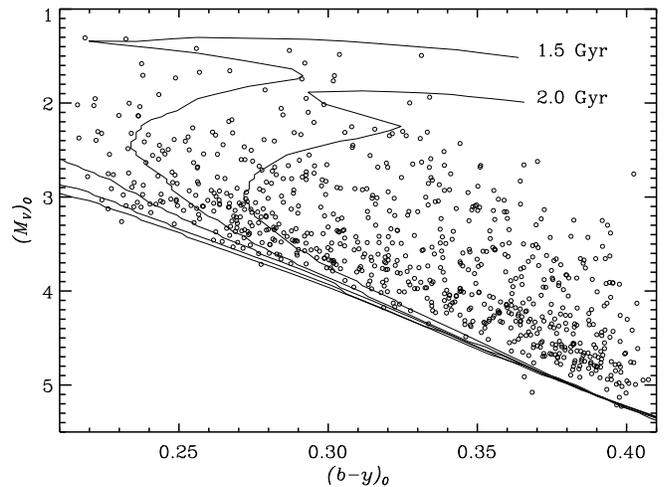}
\caption[CM diagram of Field stars with isochrones]
{\label{isoplot}
{Color-magnitude diagram of the selected field stars with 
$-0.10<[\mathrm{Fe/H}]<0.12$ and $\sigma (\pi)/\pi<0.05$, 
compared to five isochrones representing the ages: 100 Myr, 500
Myr, 1 Gyr, 1.5 Gyr, and 2 Gyr.
}}
\end{figure}

Figure \ref{isoplot} shows the field stars in the color-magnitude
diagram and five isochrones of ages: 100 Myr, 500 Myr, 1 Gyr, 1.5 Gyr,
and 2 Gyr. 
It is evident
that the sample of field stars contains many stars with ages
below or around 1.5 Gyr, which are all the stars at the left side of or
around the 1.5 Gyr
isochrone. Since we safely can assume that the age of the field
stars is distributed uniformly in the plotted color range there must
also be many field stars of ages around or less than 1.5 Gyr at the right side
of the plot say in the range $0.30<(b-y)_{0}<0.40$. 
There may be a systematic offset of the isochrones with respect to
the ZAMS$_{\mathrm{Field}}$ due to systematic errors in the $T_{\mathrm{eff}}$
calibration, but Fig. \ref{isoplot} shows that the evolutionary
effects on the isochrones from 100 Myr (approximate age of the
Pleiades cluster) to 1.5 Gyr is rather negligible in
the color range $0.30<(b-y)_{0}<0.40$. In that range we still see the
discrepancy between the position of the Pleiades and the
ZAMS$_{\mathrm{Field}}$ if the Hipparcos distance is adopted. Hence, we
conclude that the discrepancy cannot be explained as due to evolution
away from the ZAMS of the field star population.

As an additional check, evolutionary tracks from Lejeune \&
Schaerer (\cite{lejeune}) have been used to calculate the
theoretical stellar magnitude distribution
at a given $(b-y)_{0}$ in order to see how well it agrees with the
underlying evolutionary distribution of $M_{V}$ adopted in
Sect. \ref{star_dens_profil}. We assume a constant star formation
rate for solar metallicity stars over the lifetime of the galactic
disk ($\sim 8$ Gyr) in agreement with the age-metallicity diagram
(Fig. 14) of Edvardsson et al. (\cite{edvardsson1993}), and a
constant initial mass function over the small mass range
corresponding to a given
$(b-y)_{0}$. The calculated magnitude distribution is similar to
the underlying evolutionary distribution of $M_{V}$ adopted in Sect. 
\ref{star_dens_profil}, i.e. with a sharp
edge at the ZAMS and a steep evolutionary fall-off. The
fall-off is less steep at the blue end of the $(b-y)_{0}$ range and
somewhat steeper at the red end, and the fall-off has a
tendency of a two-step function; first a steeper part followed by a
less steep part. As discussed in
Sect. \ref{star_dens_profil} on page \pageref{two-step} this could
indicate that our fitted Pleiades distance modulus is slightly
underestimated. Altogether, we conclude that the
assumed underlying evolutionary magnitude distribution of our field
stars is supported by models for the stellar evolution.

\subsection{[Fe/H] sensitivity of ZAMS$_{Field}$}\label{metalsens}
It is known that the ZAMS locus is dependent on the 
metallicity, such that low metallicity stars define a 
fainter ZAMS than the high metallicity stars. 
To be able to test if the discrepancies in the 
Pleiades distance could be explained 
as a possible error in the adopted Pleiades 
metallicity, we have analyzed how much the locus 
of the ZAMS$_{\mathrm{Field}}$ changes as a function of the 
metallicity. 

The selected sample for this part of the investigation 
consists of all F-type field 
stars in the catalogue of Olsen (\cite{olsen99}), which have a
relative error in the parallax measurement less than 5\%. This
selection gives a sample of 2309 stars. 
Five plots, each representing field stars in different 
metallicity intervals, are then made. The intervals 
are $\pm 0.10$ dex wide in [Fe/H], ranging from $-0.45$ 
to $+0.15$ dex, and 
with a 0.10 dex overlap from one interval to the next.

\begin{figure}\centering
\includegraphics{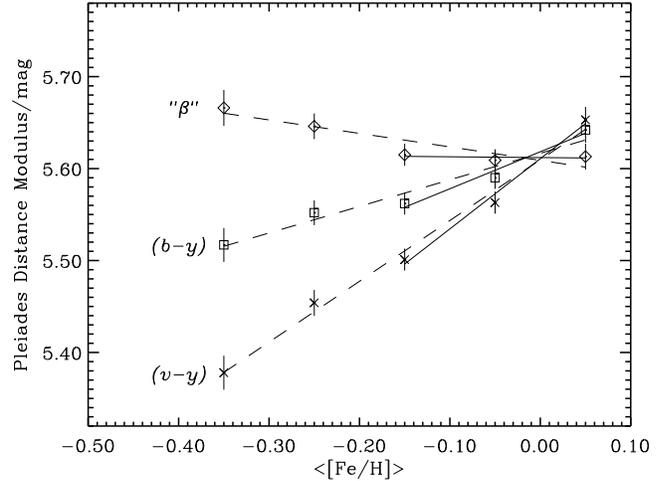}
\caption[Metallicity sensitivity of three CM diagrams]
{\label{metal_sens_plot}
{Metallicity sensitivity of the ZAMS$_{\mathrm{Field}}$ in the three
investigated color-magnitude diagrams: $\beta$ (diamonds),
$(b-y)$ (squares), and $(v-y)$ (X's). Solid and dashed lines
are least squares fits (see text).
}}
\end{figure}

Distance moduli for the Pleiades are found for the five metallicity
intervals in all three color-magnitude diagrams. The method is similar
to that presented in Sect. \ref{star_dens_profil}, and the 
individual distance moduli from every diagram is plotted together in
Fig. \ref{metal_sens_plot}. The indicated error bars in
Fig. \ref{metal_sens_plot} are
larger for the data points representing the lower metallicity
intervals, which is a result of less stars falling in the lower
metallicity bins. By analyzing the fits to the field star density
profile of
the lower metallicity intervals it was observed that the fitting
function (see Sec \ref{star_dens_profil} Eq. (\ref{convolution})) did
not fit the left tail of the distribution
as good as shown in Fig. \ref{dens_profile} simply because of the few
data points available.
Linear least squares fits are made to the result of all three
color-magnitude diagrams in the metallicity range
$-0.35<[\mathrm{Fe/H}]<0.05$ (dashed lines), and additional fits where the two
lowest metallicity points are ignored (solid lines).

The metallicity sensitivity from the three color-magnitude diagrams is
found to be (solid lines):
\[
\begin{array}{lp{0.8\linewidth}}
 \Delta \left <M_{V_{0}}\right >_{\mathrm{ZAMS}}=
 +0.01(9)\, \Delta[\rm{Fe/H}],    & from $\beta$      \\
 \Delta \left <M_{V_{0}}\right >_{\mathrm{ZAMS}}=
 -0.40(9)\, \Delta[\rm{Fe/H}],    & $(b-y)_{0}$       \\
 \Delta \left <M_{V_{0}}\right >_{\mathrm{ZAMS}}=
 -0.76(9)\, \Delta[\rm{Fe/H}],    & and $(v-y)_{0}$.  \\
\end{array}
\]

The trend is the same for both solid and dashed lines, though solid
lines show a higher metallicity sensitivity in $(b-y)_{0}$ and
$(v-y)_{0}$ and no significant sensitivity in the $\beta$ diagram.
We believe that the relations indicated by the solid lines are the most
correct since the lower metallicity bins may suffer from small 
evolutionary effects due to a larger percentage of more evolved stars. 
That the highest metallicity sensitivity is observed in the $(v-y)_{0}$
diagram and the
lowest in the $\beta$ diagram is expected since line blanketing from
the metal lines
affects the spectrum more at the short wavelengths, thus $v$ is more
affected than $b$ which is again more affected than $y$. 
We note that Grenon (\cite{grenon1999})
finds a much larger shift 
$\Delta \left<M_{V_{0}}\right>_{\mathrm{ZAMS}}=-1.67\, \Delta[\rm{Fe/H}]$
based on Geneva photometry, though the error of the metallicity
sensitivity coefficient is not given. His relation refers, however, to early
K-dwarfs and it is not clear on which color index it is
based. A direct comparison can therefore not be made to the shifts we
are finding for the F-type stars.

Figure \ref{metal_sens_plot} supports that the metallicity
of the Pleiades is around the solar value as determined from
the $m_{1}$ index if the adopted Hyades metallicity is $\sim 0.12$
dex. Only for this metallicity can we 
get consistent results of the distance modulus from all three
color-magnitude diagrams.
The inferred distance modulus is in the range 5.58-5.64 mag. We also
see that a Hipparcos distance modulus of 5.36 mag is not consistent
with one metallicity value. The $(v-y)_{0}$ diagram demands
$[\mathrm{Fe/H}]\sim -0.35$,
the $(b-y)_{0}$ diagram $[\mathrm{Fe/H}]\sim -0.70$, and the $\beta$ diagram
can not be fitted to a distance modulus of 5.36 mag no matter the
chosen metallicity of the field stars. 

From the above metallicity sensitivity analysis one could conclude
that the $\beta$ diagram gives the most reliable estimate of the 
distance modulus, because it
is rather insensitive to metallicity deviations between field stars
and Pleiades stars, and furthermore it is insensitive to
interstellar reddening.
It should, however, be remembered that the $\beta$ diagram has the
largest error in the distance modulus determination
(0.03 mag); 3/2 the size of the $(b-y)_{0}$ and $(v-y)_{0}$ diagram errors. If we
combine the distance moduli from all
three color-magnitude diagrams the final distance modulus is:
$(m-M)_{0}=5.61\pm0.03$ mag\footnote{The 
distance modulus found by CP76, based on all 30 member 
stars in their Table II, was $(m-M)_{0}=5.53\pm0.04$ mag, and 
by rejecting four stars, probably double stars or non-members, their 
result was $(m-M)_{0}=5.60\pm0.03$.}.

\section{Discussion and conclusion}\label{discussion}
The extensive multi color MS fitting analysis based on Str\"omgren
photometry (this paper) gives a Pleiades distance modulus of
$5.61\pm0.03$ mag (the
mean value from all three color-magnitude diagrams), which is in very
good agreement with the distance modulus given by the multi color MS
fitting analysis of Pinsonneault et al. (\cite{Pin98}) who find a
distance modulus of $5.60\pm0.05$ mag.
On the other hand, the 
distance modulus derived from the Hipparcos parallaxes 
are all in the range of 5.33-5.37 mag, which is not 
consistent with the former results, and the quoted errors. 
Even if the quoted errors from MS fitting are not 
representative for the actual uncertainty in this method, 
it must, from the current 
investigation, be concluded that the Hipparcos distance 
modulus is not consistent with the ZAMS of nearby field 
stars (Fig. \ref{plei_hip_dist}). Thus, either are the 
Hipparcos parallaxes affected by systematic errors, which 
are significantly larger than expected, or there is 
something unexpected about the Pleiades cluster. 

If the discrepancy is caused by some anomaly of the 
Pleiades, one possible explanation is the metallicity. 
Extensive 
investigation of this possibility has been performed 
(e.g. Pinsonneault et al. \cite{Pin98} and the current investigation
presented in Sect. 
\ref{metalsens}), and from these results, supported by the
spectroscopic metallicity determination by Boesgaard \& Friel
(\cite{BoesFriel}), it seems very 
unlikely that the adopted Pleiades metallicity around the 
solar value is more than 0.05 dex from the true value. Even a 
deviation of 0.1 dex is far too small to explain 
the discrepancy and it would lead to inconsistent results from the
different color-magnitude diagrams. 
In addition, the relative comparison 
between the Pleiades and field stars of the same metallicity 
(Sect. \ref{min_del}) ensures that the possible difference in 
the relative 
metallicities are so small that a metallicity 
deviation can be ruled out. Hence, we cannot confirm the recent claim
by Grenon (\cite{grenon1999}) that the Pleiades problem is solved by
adopting a low metallicity, $[\mathrm{Fe/H}]_{\mathrm{Pleiades}}=-0.11$, of the
cluster. Furthermore, the large metallicity difference between the
Hyades and the Pleiades based on Geneva photometry,
$\Delta[\mathrm{Fe/H}]=0.25\pm0.03$, is inconsistent with the difference,
$\Delta[\mathrm{Fe/H}]=0.11\pm0.03$, which we derive from Str\"omgren
photometry.

Another possibility is an abnormal helium abundance of the 
Pleiades. To see which value of $Y$ would be required to change 
the Pleiades MS locus by 0.3 mag, a calculation is made 
by Pinsonneault et al. (\cite{Pin98}), who find that the value is
as high as $Y\simeq0.37$.
The study of Nissen (\cite{nissen1974}) revealed no intrinsic
scatter in $Y$ greater than approximately 10\% in 
nearby MS field B stars; much smaller than the 30\%-40\% change in $Y$
required for the Pleiades.
There are, however, investigations which indicate large 
cluster to cluster scatter in the helium abundance (Nissen
\cite{nissen1976}; Lyubimkov \cite{Lyubimkov1977}), and it has been
suggested that this is the key explanation to 
the Hyades $c_{1}$-anomaly (Str\"omgren et al. \cite{stromgren}). To
test if this is the explanation of the 
Pleiades problem too, an attempt should be made to measure the surface
helium abundance of the hot stars in the Pleiades
and other young clusters spectroscopically. 

Recently, van Leeuwen (\cite{vanleeuwen99}) has suggested that the Pleiades 
problem is caused by an age effect, and claims that other very young
open clusters show the same deviation as the Pleiades. 
The investigation of van Leeuwen 
(\cite{vanleeuwen99}) is based on a comparison in the color-magnitude 
diagram (with the metallicity-sensitive $B-V$ color) of nine 
open clusters, all shifted to their Hipparcos mean distance. 
But this is done without correcting for differences in the 
metallicity abundances first. 
A test of the age effect is made by Pinsonneault 
et al. (\cite{Pin99}), based on 8 clusters (incl. 
the Hyades and Pleiades) and no age effect 
is seen in the difference between the MS fitting and 
Hipparcos distances. 
If the youth of the Pleiades has a significant 
affect on the Pleiades MS locus, one might also expect that young
field stars would show the same effect.
The study of Soderblom et al. (\cite{Sod98}) of chromospherically
active (and therefore assumed young) stars gave, however, no
indications of that. In our own sample of field stars very few are
expected to be as young as the Pleiades so we cannot test the
suggestion of van Leeuwen (\cite{vanleeuwen99}). It should be
emphasized, however, that the explanation given by van
Leeuwen (\cite{vanleeuwen99}) of the Pleiades
problem as an age effect, mostly relies on the relative shift between
the Pleiades and Hyades in the temperature range where no
obvious evolutionary effects away from the ZAMS are seen (corresponding to
$(b-y)_{0}>0.30$ mag or $(B-V)_{0}>0.50$ mag). Furthermore,
the age explanation of van Leeuwen (\cite{vanleeuwen99}) is in
disagreement with theoretical models of stellar
evolution, which predict negligible evolution away from the ZAMS
during the first couple of billion years of the lifetime of late
F-type stars (see isochrones in Fig. \ref{isoplot}). 

Some investigations give hints of spatial systematic errors 
in the Hipparcos parallaxes which are larger than expected 
(Pinsonneault et al. \cite{Pin98}; Narayanan \& Gould
\cite{narayanan99}). 
In addition, there is a statistical 
correlation caused by the imperfect distribution 
of data points over the ellipse described by the 
parallactic motion. In particular, for a star on the 
ecliptic, equal numbers of measurements should be 
obtained on both sides of the Sun. This was not 
fulfilled for Hipparcos, and caused correlations ($\rho^{\pi}_{\alpha}$)
between right ascension and parallax\label{pin98suncorr} 
(ESA \cite{esa1997}, Vol.1, p.325). In Pinsonneault et
al. (\cite{Pin98}) it is questioned if this type of correlation would
have an effect on the parallax values. But from the extensive test by
Robichon et al. (\cite{Rob99})
of this issue it must be concluded that the correlations
$\rho^{\pi}_{\alpha}$ do not introduce significant errors in the Hipparcos
parallaxes. 

What if the whole controversy, about the Pleiades 
distance modulus, is caused by a real effect? What will the 
effects on 
the MS fitting and Hipparcos results be, if the cluster 
is non-spherical (sphericity has until now been implicitly 
assumed)? 
The angular size of the Pleiades cluster can be approximated 
by the size of the region covered by the member stars used 
in the 
investigation of e.g. Narayanan \& Gould
(\cite{narayanan99}). This gives 
a radius of $\sim 6^{\circ}$ which, at a distance 
of 130 pc, corresponds to a radius of $\simeq 14$ pc. A 
typical real difference in the cluster member distances 
would then be around 14 pc 
which corresponds to a 1 mas difference in the parallax. 
There is no reason why non-sphericity of the Pleiades 
should not be the case. There is plenty of evidence that 
open clusters can be non-spherical e.g NGC 2264 
(the Cone Nebulae), and actually Raboud \& Mermilliod
(\cite{raboud1998}) have shown that the distribution of the Pleiades
stars projected on the sky is elliptical with an ellipticity of
0.17. It could be that the cluster has a more oblong shape in the
direction of the line of sight say with a length that is twice the
projected diameter.
One could then imagine the following scenario: The first born 
bright stars (O and B-type) forms in one part of the gas 
cloud, and they start to 
blow the gas cloud in one initial direction, and therefore 
these stars will end up at one end of this 
deform shape (as observed in the NGC 2264 case), and 
the fainter stars (F and G-type) will form a ``tail'' (as 
an overall trend). So if we see this shape head-on there 
will be a trend that the brightest B-type stars are closer to us, 
and the later classes are further away. 
Because the calculation of the Hipparcos mean parallax 
gives the largest weight to the brighter stars, the 
result will be a slightly shorter distance, than the actual 
mean cluster distance (Pinsonneault et al. \cite{Pin98}, Fig. 20).
On the other hand, the MS fitting 
method relies mostly on the fainter stars (A to G-type), 
which are farther away. 
These stars are located in the color-magnitude diagram, where 
the slope of the ZAMS is less steep, and therefore they give the 
smallest errors in the distance estimates (additionally for many 
clusters the hotter stars are also evolved away from the ZAMS). 
So this means 
that the distance found from MS fitting will be larger 
than found from Hipparcos parallaxes, and possibly closer 
to the actual mean cluster distance. 
Such oblong shape of the cluster, could 
indicate a kinematic history that does not follow 
the usual assumptions, 
which also explains why Robichon et al. (\cite{Rob99}) 
find unusual features in the kinematics of the Pleiades. 
The consequence of this is that the assumptions used 
by Narayanan \& Gould
(\cite{narayanan99}) does not hold. 
One could argue that if the Pleiades have such non-spherical 
shape, it would be expected that some fainter stars 
(which have larger distances) were present near the center 
of the cluster in e.g. Fig. 20 of Pinsonneault et
al. (\cite{Pin98}). But if there 
is a dark cloud just behind the bright stars as in the 
NGC 2264 case, none of these stars will be observed. 

The idea of a deform Pleiades cluster is a
tempting answer to the Pleiades problem, because 
it includes most of the evidence presented in this 
discussion. 
What is described above, as a possible non-spherical Pleiades 
cluster, shall be viewed as an illustration or example 
of the possible effects on the distance determination, due 
to a deform and non-symmetrical cluster. 
Though there are studies of the velocity dispersion among
Pleiades cluster members (van Leeuwen \cite{vanleeuwen94}) and mass
segregation (Raboud \& Mermilliod \cite{raboud1998}) suggesting that
the Pleiades is a bound and quite relaxed system, it could be very 
interesting to investigate the possibility of a non-symmetrical cluster
by e.g. extensive kinematic analysis of the Pleiades.
Furthermore, future astrometric space programs will be capable of 
determining the distances to the individual Pleiades stars 
with an improved accuracy of 2 to 3 orders of magnitude, 
compared with Hipparcos. These measurements will provide 
a very good three-dimensional picture of the Pleiades cluster. 

\begin{acknowledgements}
We thank F. Grundahl for the merging of The Hipparcos Catalogue 
and the field stars catalogue, used in this investigation. This
research has made use of the SIMBAD database, operated at CDS,
Strasbourg, France.
\end{acknowledgements}

\end{document}